\documentclass[useAMS,usenatbib]{mn2e}
\usepackage{graphicx}
\usepackage{txfonts}
\title[Neutron star-Be star's disc interaction]{On the neutron star-disc interaction in Be/X-ray binaries}
\author[P. Reig]{P. Reig$^{1,2}$\thanks{E-mail: pau@physics.uoc.gr} 
\\
$^{1}$IESL, Foundation for Research and Technology, 71110 Heraklion,
Crete, Greece\\
$^{2}$University of Crete, Physics Department, PO Box 2208, 710 03
Heraklion, Crete, Greece}

\newcommand{\bex}    {BeX}

\def\simless{\mathbin{\lower 3pt\hbox
     {$\rlap{\raise 5pt\hbox{$\char'074$}}\mathchar"7218$}}}   %< or of order
\def\simmore{\mathbin{\lower 3pt\hbox
     {$\rlap{\raise 5pt\hbox{$\char'076$}}\mathchar"7218$}}}   %> or of order

\begin{document}

\date{Accepted ??. Received ??; in original form ??}

\pagerange{\pageref{firstpage}--\pageref{lastpage}} \pubyear{2007}

\maketitle

\label{firstpage}

\begin{abstract}
We have investigated the long-term X-ray variability, defined as the
root-mean-square ($rms$) of the ASM RXTE light curves, of a set of galactic
Be/X-ray binaries and searched for correlations with system parameters,
such as the spin period of the neutron star and the orbital period and
eccentricity of the binary. We find that systems with larger $rms$
are those harbouring fast rotating neutron stars, low eccentric and
narrow orbits. These relationships can be explained as the result
of the truncation of the circumstellar disc. We also present an updated version of the
H$\alpha$ equivalent width--orbital period diagram, including sources in the
SMC. This diagram provides strong observational evidence of the interaction
of neutron star with the circumstellar envelope of its massive companion. 
\end{abstract}

\begin{keywords}
X-rays: binaries -- stars: neutron -- stars: binaries close --stars: 
 emission line, Be
\end{keywords}

\section{Introduction}

Be stars are non-supergiant fast-rotating B-type and luminosity class III-V
stars which at some point of their lives have shown spectral lines in
emission. In the infrared they are brighter than their non-H$\alpha$
emitting counterparts of the same spectral type. The line emission and IR
excess originate in extended circumstellar envelopes of ionised gas
surrounding the equator of the B star.   Be stars can live their lives in
isolation or taking part in a binary system. In this case, the companion is
a neutron star and the system is referred to as a Be/X-ray binary
\citep[see][for recent reviews]{coe00,negu04,negu05}. Evolutionary calculations
show that Be star + white dwarf or Be star + black hole should also be
common type of systems. However, no clear evidence of their existence has
been shown as yet \citep{wate89,ragu01,torr01,zhan04}.  Be/X-ray binaries
(from now on \bex) are strong emitters of X-ray radiation, which is
produced as the result of accretion of matter from the optical companion's
circumstellar disc onto the neutron star. 

The variability time scales in \bex\ range from seconds to years. The
fastest variability is found in the X-ray band. All but two (LS I +61303
and 4U 2206+54) of the \bex\ with identified optical counterparts are X-ray
pulsars. Pulse periods cover the range 3.6--1412 s. On longer time scales
(months to years), the variability is also apparent in the optical and IR
bands and it is attributed to structural changes of the circumstellar
disc.  Sometimes, the Be star loses the disc
\citep{roch97,haig99,negu01,reig01}. When this occurs the H$\alpha$ line
shows an absorption profile and the X-ray activity ceases. The long-term
X-ray variability of the transient \bex\ is characterised by two type of
outbursting activity 

\begin{itemize}

\item Type I outbursts. These are regular and (quasi)periodic outbursts,
normally peaking at or close to periastron passage of the neutron star.
They are short-lived, i.e., tend to cover a relatively small fraction of
the orbital period (typically 0.2-0.3 $P_{\rm orb}$). The X-ray flux
increases by up to two orders of magnitude with respect to the pre-outburst
state, reaching peak luminosities $L_x \leq 10^{37}$ erg s$^{-1}$.

\item Type II outbursts represent major increases of the X-ray flux,
$10^{3}-10^{4}$ times that at quiescence. They reach the Eddington
luminosity for a neutron star ($L_x \sim 10^{38}$ erg s$^{-1}$) and become
the brightest objects of the X-ray sky. They do not show any preferred
orbital phase and last for a large fraction of an orbital period or even
for several orbital periods. The formation of an accretion disc during Type
II outbursts  \citep{kris83,motc91,haya04} may occur. The discovery of
quasi-periodic oscillations in some systems \citep{ange89,fing96} would
support this scenario. The presence of an accretion disk also helps explain
the large and steady spin-up rates seen during the giant outbursts, which
are difficult to account for by means of direct accretion.

\end{itemize}

Most \bex\ are transient systems and present moderately eccentric orbits
($e\simmore0.3$), although persistent sources also exist. Persistent \bex\
\citep{reig99} display much less X-ray variability and lower flux ($L_x
\simless 10 ^{35}$ erg s$^{-1}$) and contain slowly rotating neutron stars 
($P_{\rm spin} > 10^2$ s). \citet{pfah02} suggested the existence of a new
class of \bex\ (which we shall referred to as low-$e$ \bex), characterised
by wide orbits ($P_{\rm orb}> 30$ d) and low eccentricity ($e\simless
0.2$). The interest  in these sources resides in that their low
eccentricity requires that the neutron star received a much lower kick
velocity at birth than previously assumed by current evolutionary models.

This work investigates the relationship between the long-term X-ray/optical
variability and the orbital parameters of galactic \bex. The parameter that
characterises the X-ray variability is the root-mean-square calculated from
the ASM RXTE light curves. In the optical band, the maximum value ever
reported of the equivalent width of the H$\alpha$ line serves as an
indication of the size of the circumstellar disc. These quantities are
studied as a function of the size (as given by the orbital period) and
eccentricity of the orbit and the rotation velocity of the neutron star
(as given by the spin period). The ultimate goal is to provide further
evidence of the interaction between the neutron star and the Be star's
envelope. 
\begin{table*}
\centering
% \begin{minipage}{140mm}
\caption{X-ray properties of the galactic Be/X-ray binaries of our sample.}
\label{sources}
\begin{tabular}{lcccccccc}
\hline \hline
Source	 &Spectral&P$_{\rm spin}$ &P$_{\rm orb}$ &$e$  &Average$^{a,c}$ &$\sigma_X^{b,c}$   &Peak$^{a,c}$ &Outburst  \\
name 	&type		&(s)   	&(days)	&	    &Flux    		& 		&Flux		&activity \\
\hline
\multicolumn{9}{c}{Transient \bex} \\
\hline
4U 0115+634     &B0.2V        &   3.6  & 24.3	&0.34 &0.49   & 2.19   &16.36 &II \\
V 0332+53       &O8-9V        &   4.4  & 34.2	&0.30 &0.86   & 6.17   &62.45 &II \\
A 0535+262      &B0III        & 104.0  &111.0	&0.47 &0.24   & 0.46   & 2.52 &II  \\
4U 0726-260     &O8-9V        & 103.2  & 34.5	&--   &0.12   & 0.07   & 0.40 & I \\
RX J0812.4-3115 &B0.5III-V    &  31.9  & 81.3	&--   &0.08   & 0.11   & 0.44 & I \\
%GS 0834-430     &B0-2III-V    &  12.3  &105.8	&0.12 &0.09   & 0.08   & 0.24 & I  \\
GRO J1008-57    &O9-B1V       &  93.5  &135.0	&0.66 &0.12   & 0.09   & 0.40 & I \\
4U 1145-619     &B0.2III      & 292.4  &187.5	&$>0.5$ &0.04   & 0.20   & 0.70 & I \\
4U 1258-61	&B2V          & 272.0  &133.0	&$>0.5$ &0.11   & 0.07   & 0.30 & I \\
4U 1417-624     &B1V          &  17.6  & 42.1	&0.45 &0.16   & 0.20   & 1.54 & I  \\
%XTE J1543-568   &B0.7V        &  27.1  & 75.6	&0.03 &0.27   & 0.23   & 0.91 & I  \\
%2S 1553-542    	&--           &   9.3  & 30.6	&0.09 &0.09   & 0.24   & 1.48 & I \\
GS 1845-024     &--           &  94.8  &241.0	&0.88 &0.18   & 0.10   & 0.36 & I  \\
XTE J1942+274   &B0-1         &  15.8  &169.2	&0.33 &0.31   & 0.56   & 1.73 & I \\
%KS 1947+300     &B0V          &  18.7  & 40.4	&0.03 &0.48   & 1.02   & 6.78 &I+II   \\
EXO 2030+375    &B0.5III-V    &  41.8  & 46.0	&0.41 &1.30   & 4.31   &28.86 &I+II   \\
GRO J2058+42    &O9.5-B0IV-V  & 198.0  & 55.0	&--   &0.15   & 0.10   & 0.46 & I  \\
SAX J2103.5+4545&B0V          & 358.6  & 12.7	&0.40 &0.28   & 0.33   & 1.59 & I  \\
%SAX J2239.3+611	&B0-2         &1247.0  &262.0   &0.00 &0.02   & 0.02   & 0.06 & & \\
\hline
\multicolumn{9}{c}{Persistent \bex} \\
\hline
RX J0146.9+6121	&B1III-V      &1412    &300$^d$	&--   &0.08   &0.10	&0.38	&-- \\
X-Per		&O9.5III      & 837    & 250	&0.11 &1.28   &0.62	& 2.74	&--\\
RX J0440.9+4431	&B0.2V	      &202     &150$^d$	&--   &0.04   &0.04	&0.18	&--\\
RX J1037.5-5648	&B0V-III      &862     &250$^d$	&--   &0.05   &0.04	&0.12	&--\\
\hline
\multicolumn{9}{c}{Low-$e$ \bex} \\
\hline
GS 0834-430     &B0-2III-V    &  12.3  &105.8	&0.12 &0.09   & 0.08   & 0.24 & I  \\
XTE J1543-568   &B0.7V        &  27.1  & 75.6	&$<0.03$ &0.27   & 0.23   & 0.91 & I  \\
2S 1553-542    	&--           &   9.3  & 30.6	&$<0.09$ &0.09   & 0.24   & 1.48 & I \\
KS 1947+300     &B0V          &  18.7  & 40.4	&0.03 &0.48   & 1.02   & 6.78 &I+II   \\
\hline \hline
\multicolumn{9}{l}{$a:$ In ASM RXTE c/s, for the period MJD 50133-54041. Note that the Crab nebula flux is about 75 ASM c/s}\\
\multicolumn{9}{l}{$b:$ Standard deviation of the ASM light curve between MJD 50133-54041} \\
\multicolumn{9}{l}{$c:$ Bin time equal to the orbital period}\\
\multicolumn{9}{l}{$d:$ Orbital period estimated from the $P_{\rm spin}$/$P_{\rm orb}$ relationship}\\
\end{tabular}
%\end{minipage}
\end{table*} 
%------------------------------------------------------------------------------

%------------------------------------------------------------------------------
\begin{figure*}
\includegraphics[width=16cm]{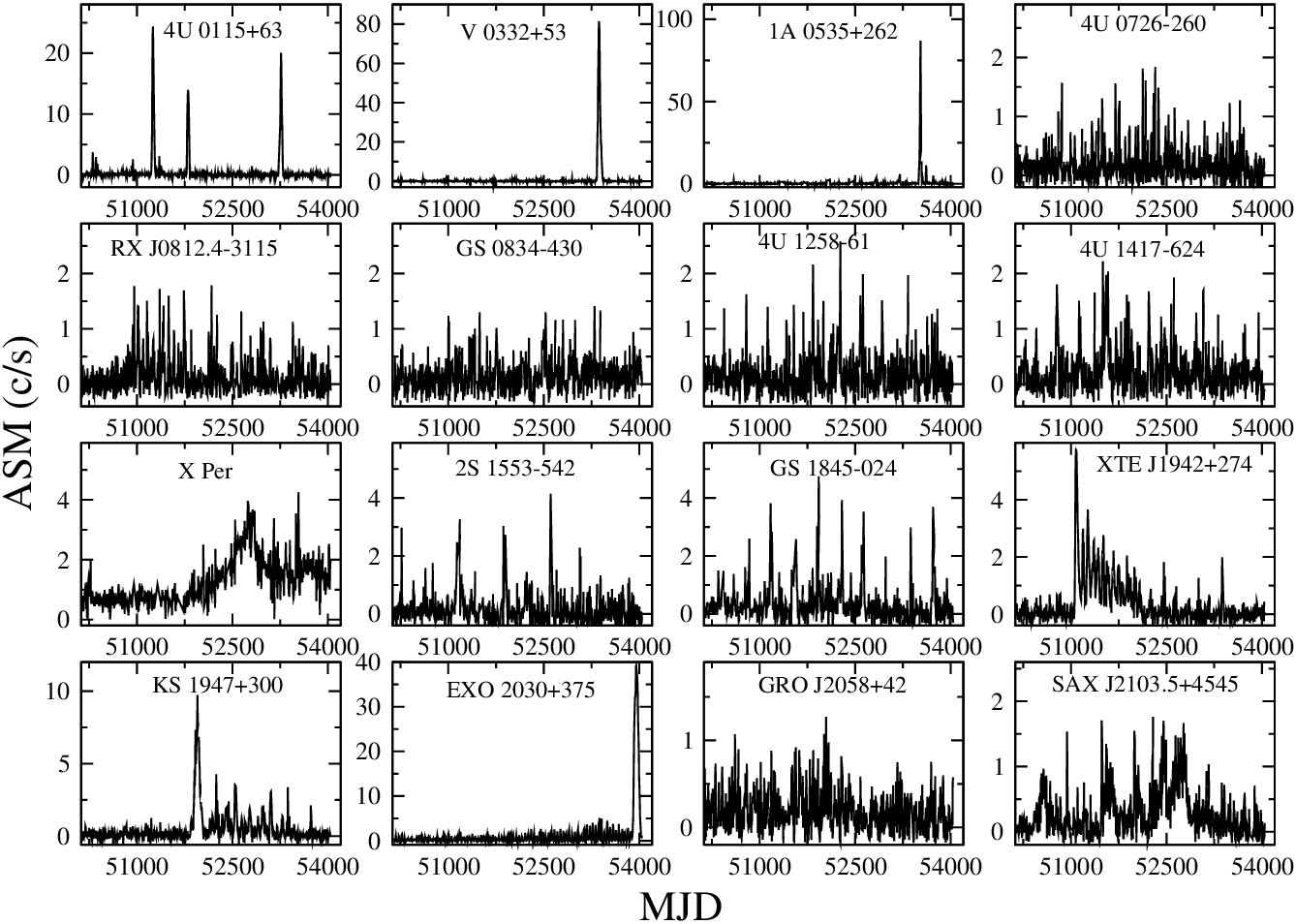}
\caption[]{ASM light curves of some of the Be/X-ray binaries investigated
in this work. The light curves cover the interval from February 1996 
to November 2006. The various types of X-ray
activity are apparent. Bin size equals 5 days.}
\label{lc}
\end{figure*}
%------------------------------------------------------------------------------
%------------------------------------------------------------------------------
\begin{figure*}
\begin{center}
\includegraphics[width=16cm]{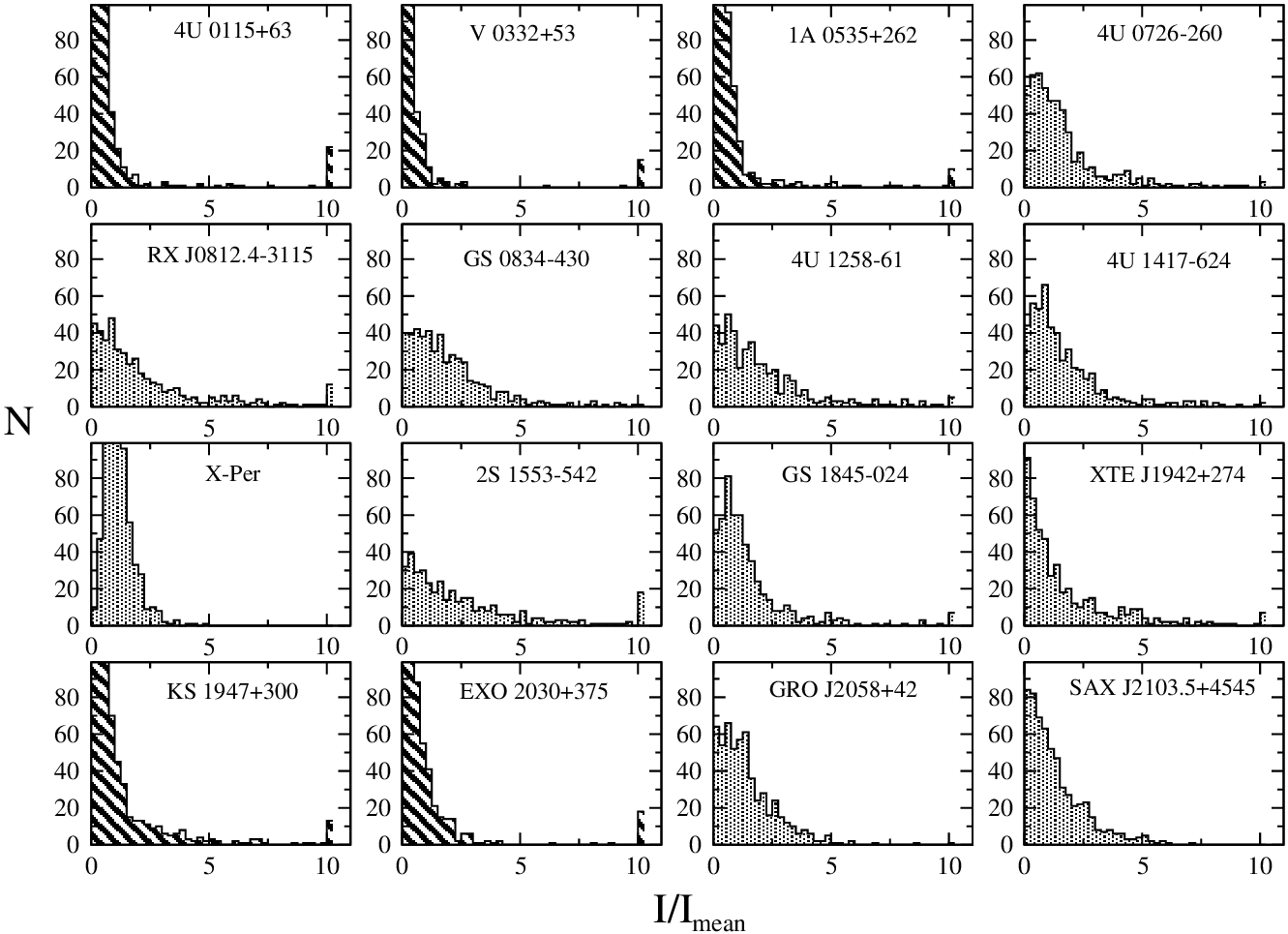}
\caption[]{Histograms of the individual sources. Sources displaying type II
outburst present narrow IDFs and have been marked with striped shaded
areas. 5-day rebinned light curves were employed.}
\label{idf}
\end{center}
\end{figure*}
%------------------------------------------------------------------------------

\section{Observations and data reduction}

Light curves obtained with the {\em All Sky Monitor} (ASM) on board the
{\it Rossi X-ray Timing Explorer} (RXTE) were retrieved from the Definitive
Products Database. The  light curves cover more than 10 years worth of
data, from February 1996  to November 2006 (JD 2,450,133--2,454,041).   The
ASM consists of three Scanning Shadow Cameras mounted on a rotating drive
assembly. Each Camera has a field of view (FWHM) of $6^{\circ} \times
90^{\circ}$ and is equipped with position-sensitive Xenon proportional
counters with a total collecting area of 90 cm$^2$. Two of the cameras
share the same look direction but are canted by $\pm12^{\circ}$ from each
other, while the third camera looks in a direction parallel to the ASM
drive axis. The ASM scans $\sim$ 80\% of the sky every $\sim$ 90 minutes in
a series of dwells of about 90 s each. Any given X-ray source is observed
in about 5--10 dwells every day. The ASM is sensitive to X-rays in the
energy band 1.3-12.2 keV.  For more information on the ASM see
\citet{levi96}. 

Instrumental effects were removed from the light curves by excluding the
data points that deviated more than 10$\sigma$ from the average of the five
preceding and five successive neighbouring points. Dwells with negative
flux {\em and} that deviated more than 1$\sigma$ from the average of the
total light curve were also removed. The average was calculated locally in
the case of the detection of spikes in order to prevent the removal of
outbursts, while was considered globally in the case of negative flux
observations to allow for statistical fluctuations.

Table~\ref{sources} lists the properties of the \bex\ analysed in this
work. The spin and orbital periods, eccentricity and spectral type were
taken from \citet{ragu05} and \citet{liu06}. Figure~\ref{lc} shows the
light curves of some of the \bex\ analysed in this work, while intensity
distribution functions (IDF) or histograms of the normalised flux are shown
in Fig.~\ref{idf}. The IDFs show how often a given (normalised) intensity
occurs. The abscissa represents the intensity of the source normalised to
its mean value in steps of 0.25, whereas the ordinate gives the number of
times (i.e. the frequency) that certain values of the normalised intensity
occur. Data points with a flux more than 10 times the mean flux were
allocated in the last bin. The IDF plots also give a measure of the source
variability. The five \bex\ showing type II outbursts (striped pattern in
Fig.~\ref{idf}) present narrow IDFs and a populated last bin, meaning that
most of the time they are close to quiescence and only sporadically go into
outburst. The height of the last bin indicates the strength of the
outburst, as this bin contains $I_{\rm X} > 10 \, I_{\rm mean}$. Sources
with choppy light curves will show many populated bins in their IDF plots,
hence resulting in wide IDFs or narrow IDFs with extended tails.

%The H$\alpha$ equivalent width information was compiled from the
%literature.

\section{X-ray variability}

We have examined the X-ray variability of the sources in the time domain
and considered the root mean square ($rms$) as the parameter that
characterize and quantifies that variability. Our goal is to study the
X-ray behaviour of galactic \bex\ on time scales that allow the sampling of
the dynamical evolution of the circumstellar disc ($\tau >> P_{\rm orb}$),
without the interference of possible intra-orbital variability ($\tau <
P_{\rm orb}$). Therefore, the light curves were rebinned into bins with a
duration equal to the orbital period of the source. 

The $rms$ amplitude was computed from the cleaned and rebinned light curves
as $rms=\sigma^2/\bar{x}^2$, where $\bar{x}$ is the mean count rate and
$\sigma^2=\sigma_{\rm obs}^2-\sigma_{\rm exp}^2$ is the difference between
the observed variance, $\sigma_{\rm obs}^2=\sum_{i} (x_i-\bar{x})^2/N$, and
the expected variance, $\sigma_{\rm exp}^2=\sum_{i} \sigma_i^2/N$
($\sigma_i$ are the experimental errors, and $N$ is the total number of
points). Note that the $rms$ is more sensitive to the amplitude variations than
to the frequency of those variations. In this sense, the $rms$ of a source
like, e.g. V 0332+53 will be higher than e.g.  4U 1417-624 (see
Fig.~\ref{lc}).

\begin{figure*}
\includegraphics[width=16cm]{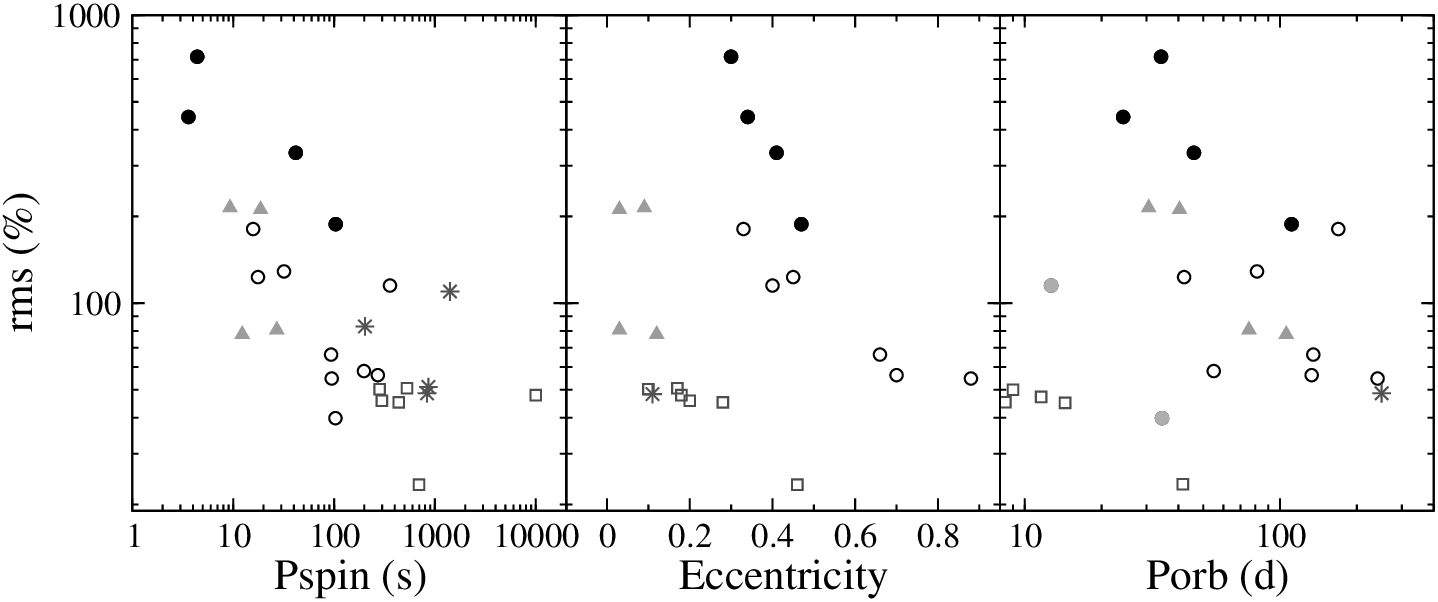}
\caption[]{$rms$ variability of the ASM light curves as a function of the
spin period (left), eccentricity (middle) and orbital period (right) of
galactic \bex. The symbols represent different type of systems as follows:
circles denote transient \bex\ systems, squares represent the
high-mass X-ray binaries with supergiant companions, triangles are the
low-$e$ \bex\ and stars correspond to persistent \bex. 
Black circles are transient \bex\ that have shown type II outbursts during 
the period covered by the observations. The two grey circles on the right 
panel correspond to 4U 0726--260 and SAX J2103.5+4545, which are \bex\ that 
do not follow the well-known  $P_{\rm spin}$/$P_{\rm orb}$ relationship.}
\label{rmsorb}
\end{figure*}

Figure \ref{rmsorb} shows the $rms$ as a function of the spin period,
eccentricity and orbital period of the systems.  For the sake of
comparison, persistent \bex\ and supergiant X-ray binaries (SGX), that is, 
systems whose optical companion is an evolved (luminosity class I-II) star,
have also been included. Different type of systems have been represented by
different symbols as follows: transient \bex\ by circles, persistent \bex\
by stars and SGX by squares. Systems that have shown Type II outbursts
during the period of the observations, that is, systems in which the
increase in X-ray flux lasted for several orbits, have been represented by
black filled circles. Grey filled triangles denote the new class of low-$e$
\bex.  It is very likely that low-$e$ \bex\ and persistent \bex\ form one
and only class of systems. In fact,  X-Per, which is considered the
prototype of the persistent \bex, also belongs to the class of low-$e$
\bex. The two grey filled circles on the right panel of Fig.~\ref{rmsorb}
correspond to 4U 0726--260 and SAX J2103.5+4545, which are \bex\ that do
not follow the well-known  $P_{\rm spin}$/$P_{\rm orb}$ relationship.

In all cases a clear anticorrelation is apparent: systems with
fast rotating neutron stars and low-eccentric and narrow orbits are
more variable, i.e. present higher $rms$. This plot then suggests two
interesting results. First, since the \bex\ and the SGX occupy clearly
distinct regions in Fig.~\ref{rmsorb}, the X-ray variability in high-mass
X-ray binaries does not only depend on the physical conditions in the
vicinity of the compact object but also on the mass transfer
mechanism, i.e. whether a circumstellar disc is present or not. Second, the
systems containing fast neutron stars are more likely to exhibit Type II
outbursts. The latter result is closely related to the inverse correlation
found by \citet{maji04} between the maximum X-ray flux and the spin period
of \bex\ in the Magellanic clouds and the Milky way. \citet{maji04}
explained this correlation in simple terms as a consequence of the inverse
dependence of the density of accreted matter (hence X-ray flux) on distance
between the Be companion and the neutron star. Since the spin period and
the orbital period are correlated through Corbet's diagram \citep{corb86},
the relationship between flux and spin period follows.

Figure \ref{rmsorb} also shows that, on average, {\em i)} the more eccentric the
orbit the lower the variability and {\em ii)} type II activity mainly occurs in
low-eccentric systems.  The newly identified class of \bex, with nearly
circular orbits (KS1947+300, XTE J1543--568, 2S 1553--542 and GS 0834-430),
represented by triangles in Fig~\ref{rmsorb}, follow the general trend in
the $rms-P_{\rm spin}$ and $rms-P_{\rm orb}$ plots but clearly distinguish
themselves from the classical \bex\ in the $rms-e$ diagram.  The
anticorrelation of the $rms$ with the orbital period provides further
evidence for the truncation of the disk in systems with short orbital
periods, i.e. those having narrow orbits.

%Since the bin size used in Fig~\ref{rmsorb} is the orbital period of each
%system, intra orbit variations are washed out.  

%The variability on time scales of days in \bex\ is determined by the
%orbital period of the system. Even during quiescent phases the X-ray flux
%have been seen to be modulated by the orbital period \citep{Coe05b}.

\begin{figure}
\begin{center}
\includegraphics[width=8cm]{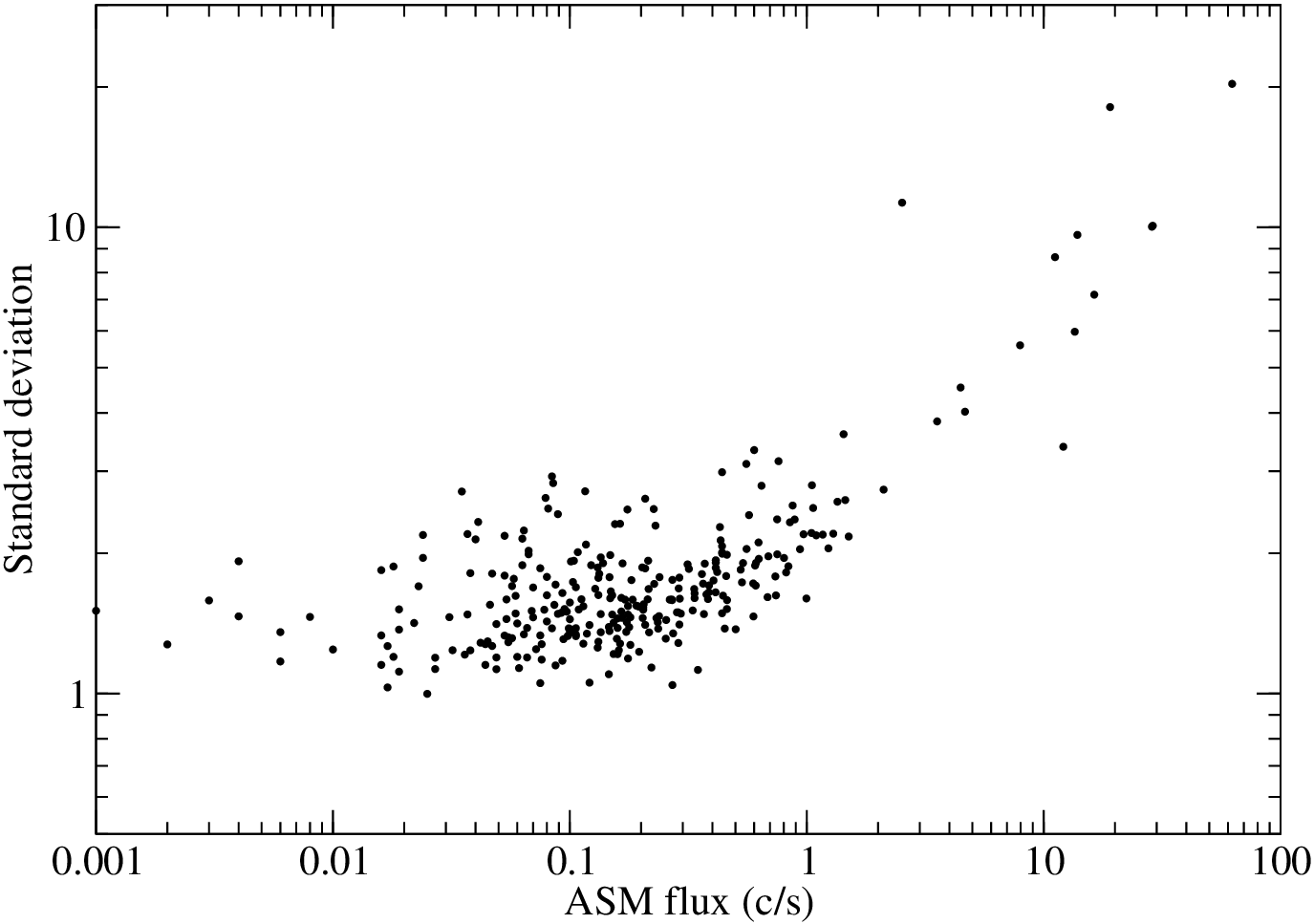}
\caption[]{Standard deviation of the ASM light curves as a function of the
flux. Only sources showing type II outbursts have been included.  }
\label{varflux}
\end{center}
\end{figure}

Systems showing large amplitude variations, that is, those exhibiting type
II outbursts, allow the study of the averaged orbital variability as a
function of X-ray flux. The ASM light curves were cut into segments of
duration equal to the orbital period. For each segment, the average count
rate and standard deviation were obtained. As can be seen in
Fig.~\ref{varflux}, the variability is consistent with statistical
fluctuations at low count rates ($<1$ ASM c s$^{-1}$), whereas at higher
intensity the X-ray variability scales with flux. The fact that there is no
accumulation of points in the bottom-right part of the diagram, indicates
that stable, long-lasting high-intensity states do not exist in \bex. High
states in \bex\ occur occasionally and are associated with short-lived and
sudden increases of the X-ray flux.

%------------------------------------------------------------------------------
%------------------------------------------------------------------------------

\section{The P$_{\rm orb}$--EW(H$\alpha$) diagram}

Figure \ref{ewpo} shows an updated version of the P$_{\rm
orb}$--EW(H$\alpha$) diagram, first presented by \citet{reig97}. The
original P$_{\rm orb}$--EW(H$\alpha$) diagram contained 11 \bex, of which
only 9 had well established orbital periods. Since then, several new \bex\
have been discovered and their orbital periods obtained. Particularly,
the discovery of new \bex\ in the Magellanic Clouds has been remarkable
\citep{stev99,edge03,maji04,coe05,layc05,liu05,schm06}. Figure \ref{ewpo}
contains 28 systems, 16 are Milky Way objects (represented by  circles) and
12 are located in the SMC (squares).  Open symbols  represent systems whose
orbital period is not yet known, but which has been estimated using an
updated version of the $P_{\rm spin}$/$P_{\rm orb}$ relationship
\citep{corb86}. Systems without a firm optical counterpart, such as AX
J0049.4--7323 (SXP756 in the terminology of \citet{coe05}), XTE J0052-723
(SXP4.78), RX J0051.8-7231 (SXP8.80) or 1WGA J0052.8-7226 (SXP46.6) or with
dubious periods (RX J0105.1-7211) have not been included in
Fig.~\ref{ewpo}. 

Note that the value of the H$\alpha$ equivalent width of Fig.~\ref{ewpo} is
the maximum ever reported. Some systems, especially in the SMC,
have been discovered recently, hence the monitoring of the H$\alpha$ is
necessarily short. For those systems, the value of the EW(H$\alpha$) may not
represent a high state of the circumstellar disc. The EW(H$\alpha$) of such systems
should be considered as lower limits (arrows pointing upward in
Fig.~\ref{ewpo} have this meaning). This is the case of the galactic \bex\
GRO J2058+42 and most likely of the SMC sources RX J0051.9-7311 and RX
J0101.3-7211.

\section{Discussion}

We have analysed the ASM RXTE light curves of galactic \bex\ to investigate
their long-term variability, with the aim of providing further evidence for
the interaction between the Be star's circumstellar disc and the neutron
star. Evidence for dynamical interaction between the neutron star and the
circumstellar envelope of the Be star was first reported by \citet{reig97},
who found a direct relationship between the maximum value of the equivalent
width of the H$\alpha$ line and the orbital period of the system. They also
found that, on average, isolated Be stars present larger values of
EW(H$\alpha$) than Be stars in \bex. These two observational facts were
interpreted as an indication that isolated Be stars may develop extended
circumstellar discs whereas the disc of Be stars in \bex\ is truncated by
the neutron star. 

Further observational support for a different disc structure in \bex\ and
isolated Be stars, presumably as a result of truncation, were subsequently
found. \citet{negu98} attributed the correlated X-ray (Type II outbursts)
and V/R optical variability to the presence of the neutron star. Based on
the works by \citet{okaz91,okaz97}, they suggests that the very short V/R
quasi-periods seen in \bex\ compared to those observed in isolated Be stars
could be explained if the former had denser circumstellar discs. This is
precisely the result found by \citet{zama01} after a comparative study
between the circumstellar discs in \bex\ and isolated Be stars.
\citet{zama01} found that the discs of Be stars in \bex\ tend to be smaller
and twice as dense as those in isolated Be stars. Likewise, the quantised
infrared excess flux states displayed by 1A 0535+262 have been interpreted
as supporting the resonant truncation paradigm \citep{haig04}. Transitions
between flux states are associated with changes in the truncation radius.
Recently, \citet{reig05} has suggested a relationship between the
characteristic time scales for the disc formation/dissipation phases and
the orbital period. That is, circumstellar discs in systems with wide
orbits last longer.

\begin{figure}
\begin{center}
\includegraphics[width=8cm]{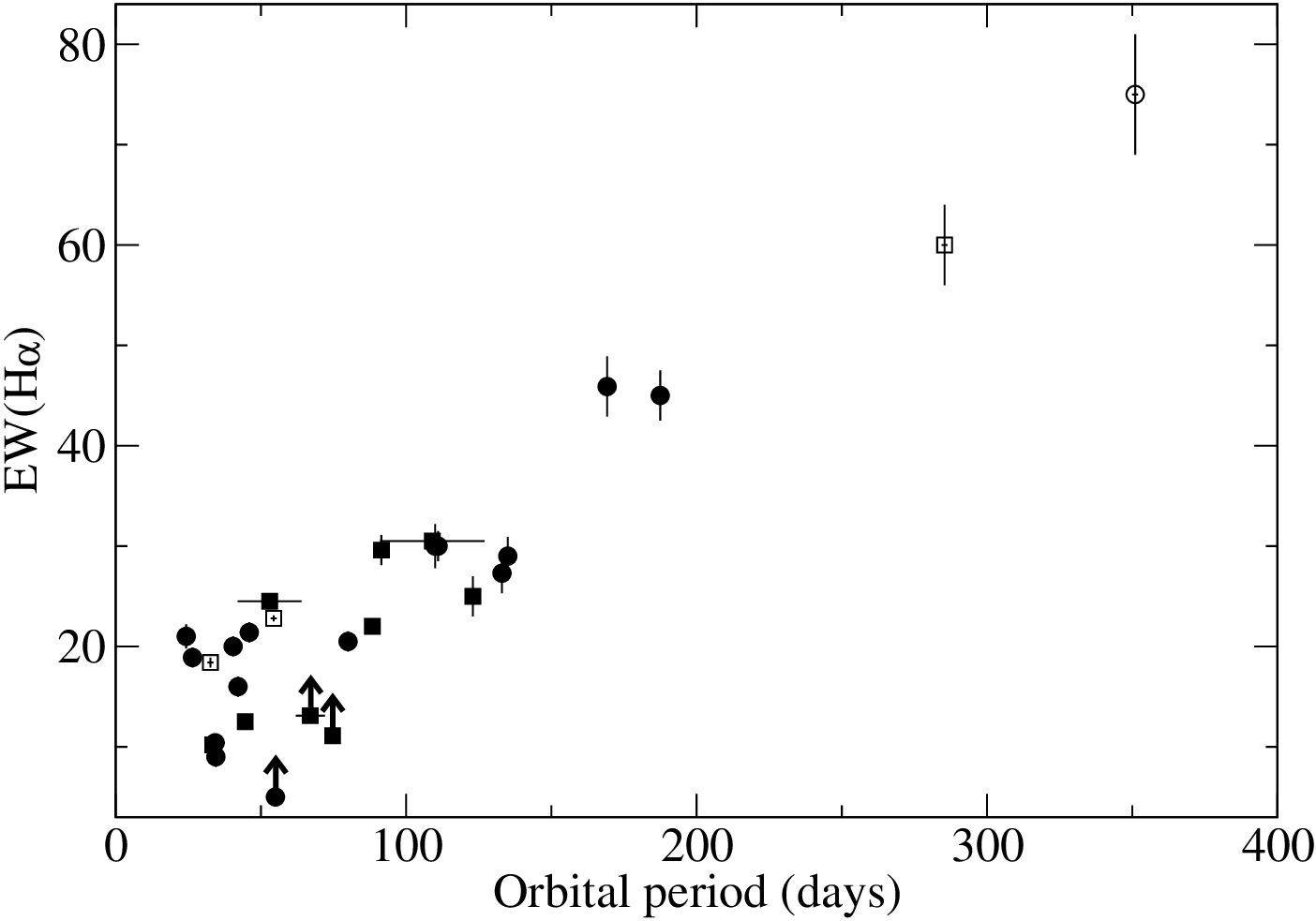}
\caption[]{P$_{\rm orb}$--EW(H$\alpha$) diagram. Milky way objects have
been represented by circles, whereas \bex\ in the Small Magellanic Cloud by 
squares. 
Open symbols represent systems whose orbital period have
been estimated using the spin-orbital period correlation.}
\label{ewpo}
\end{center}
\end{figure}

An updated version of the $EW(H\alpha)-P_{\rm orb}$ correlation is
presented in Fig.~\ref{ewpo}, where SMC sources have been included. The
P$_{\rm orb}$--EW(H$\alpha$) diagram is based on the fact that the
H$\alpha$ line is the prime indicator of the circumstellar disc state.
Although an instant measurement of the EW(H$\alpha$) may not be an
effective measurement of the size of the disc, the maximum equivalent
width, when monitored during a long length of time (longer than the typical
time scales for changes in the circumstellar disc, namely, from  few months
to a few years) becomes a significant indicator of the size of the disc. 
Based on interferometric observations, \citet{quir97} and \citet{tycn05}
have shown that there is a clear linear correlation between the net
H$\alpha$ emission and the physical extent of the emitting region. Also,
the observed correlations between the spectral parameters of the H$\alpha$
line (FWHM, EW, peak separation in double-peak profiles) and the rotational
velocity that have been observed in many Be stars are interpreted as
evidence for rotationally dominated circumstellar disc \citep{dach86}. In
particular, interpreting the peak separation ($\Delta_{\rm peak}$) of the
H$\alpha$ split profiles as the outer radius ($R_{ \rm out}$) of the
emission line forming region \citep{huan72}

\begin{equation}
\frac{R_{ \rm out}}{R_*}=\sqrt{\frac{2 v \sin i}{\Delta_{\rm peak}}} 
\label{huang}
\end{equation}

\noindent the radius of the emitting region can be estimated
\citep{humm95,jasc04}. $v\sin i$ is the projected rotational velocity of
the B star ($v$ is the equatorial rotational velocity and $i$ the
inclination toward the observer). As the EW(H$\alpha$) increases, the peak
separation decreases, hence increasing the radius of the H$\alpha$ emitting
region \citep{dach92,hanu88}. Single peak profiles tend to correspond to
larger discs. Further evidence of a direct relationship between the
EW(H$\alpha$) and the size of the circumstellar disc is the correlation
between the EW(H$\alpha$) and the infrared colours reported by various
authors \citep{dach88,coe05}. 

%------------------------------------------------------------------------------
%------------------------------------------------------------------------------

Assuming then, that EW(H$\alpha$) provides a good measure of the size of
the circumstellar disc, Fig.~\ref{ewpo} indicates that systems with long
orbital periods have larger discs, while narrow orbit systems contain
smaller discs. The existence of the above relationship is attributed to the
presence of the compact star which prevents its optical counterpart from
developing a large circumstellar disk in system with small orbital periods.
In other words, the presence of the neutron star leads to the truncation of
the circumstellar disc.

The viscous decretion disc model \citep{okaz01} provides the
theoretical basis to the idea of disc truncation. Be star's discs are
supported by viscosity. Angular momentum is transferred from the optical
companion to the inner edge of the circumstellar disc, increasing its
angular velocity to Keplerian. The radial velocity component is subsonic
even at the distance where the neutron star lies.  Truncation occurs by the
tidal interaction when the resonant torque exerted by the neutron star
exceeds the viscous torque. This occurs only at certain radii --- where the
ratio between the angular frequency of disc rotation and the angular
frequency of the mean binary motion is a rational number. The efficiency of
truncation depends strongly on the gap between the truncation
radius and the inner Lagrangian point and the viscosity parameter $\alpha$.
If $\tau_{\rm drift}$ is the time scale for a particle in the disc to cover
this gap then truncation will be efficient when $\tau_{\rm drift} > P_{\rm
orb}$. Typical values of the viscosity parameter are $\alpha < 1$.
Truncation is expected to be more efficient in low and moderate eccentric
systems ($e\simless 0.3$) with narrow orbits ($P_{\rm orb} \simless 40$ d)
than in highly eccentric systems because the former have wider gaps. The
prediction of the viscous decretion disc model is then that type I
outbursts are expected to occur more often in high-eccentric systems,
whereas type II outburst would dominate in low-eccentric systems, in
agreement with the results shown in Fig.~\ref{rmsorb}.

\citet{zhan04} extended the viscous decretion disc model to
compact companions of arbitrary mass. Their results not only confirmed the
predictions of \citet{okaz01} that low viscosity and small eccentricity
would lead to effective Be disc truncation, but also show that the most
effective truncation would occur in narrower systems, hence giving further
support to the validity of the $EW(H\alpha)-P_{\rm orb}$ correlation.

\citet{coe05} carried out a major study of the optical and infrared
characteristics of BeX in the SMC. They found that the optical photometric
variability is greatest when the circumstellar disc size is least and also in systems
containing fast rotating neutron stars. Our work extends these results to
the X-ray domain.

While the works by \citet{okaz01} and \citet{zhan04} and the
$EW(H\alpha)-P_{\rm orb}$ correlation clearly identify truncation with low
eccentric and narrow orbit systems, the plots of the {\em X-ray} $rms$ as a
function of the system parameters (Fig.~\ref{rmsorb}) and those of the 
EW(H$\alpha$) and $P_{\rm spin}$ as a function of the {\em optical}
photometric $rms$ \citep[Figs 6 and 7 in ][]{coe05}  identify truncation
with variability. That is, systems in which truncation is favoured are the
most variable ones in both, the X-ray and optical bands.

Evolutionary scenarios for the formation of \bex\ traditionally assumed
that their observed large orbital eccentricities ($e\simmore 0.2$) are the
result of substantial kick velocities ($\simmore 60$ km s$^{-1}$) imparted
on to the neutron star at birth \citep[see e.g.][]{heuv00}. The existence
of \bex\ with quasi circular orbits requires that the neutron star in these
systems received a kick velocity $\simless 50$ km s$^{-1}$
\citep{pfah02,pods04}. The large number of low-$e$ \bex\  --- they represent
more than a third of \bex\ with known orbits --- led \citet{pfah02} to
suggest that these systems conform a new class of \bex.  The high level of
X-ray activity of some members of this group, such as  KS 1947+300, which
shows both type I and II outbursts and GS 0834--430, which shows strong
type I  activity poses a problem to the viscous decretion
disc model. This model predicts that the lower the eccentricity, the more
difficult the accretion onto the neutron star as truncation is more
effective.

Although the viscous decretion disc model provides a natural explanation
for periodic X-ray outbursts in BeX, there is still a number of unresolved
issues such as the identification of the physical mechanism producing the
type II outbursts, the origin of the Be star's equatorial disc, whether an
accretion disc around the neutron star is formed and the X-ray variability
in low-$e$ systems. 

%------------------------------------------------------------------------------
%------------------------------------------------------------------------------

\section{Conclusion}

We have investigated the long-term behaviour of Be/X-ray binaries in the
X-ray and optical bands and found further evidence for the truncation of
the Be star's circumstellar disc.  The viscous decretion disc model
predicts that truncation of the Be star's disc is favoured in systems with
short orbital periods and low eccentricities. We have shown that these
systems are also the most variable in terms of flux amplitude changes
(those displaying type II outbursts). In contrast, systems with long
orbital periods and high eccentricity show lower amplitude flux variations
(type I outbursts). Our results agree and extend to the X-ray domain those
reported by \citet{coe05} in the optical band.

%The P$_{\rm orb}$--EW(H$\alpha$) diagram provides observational evidence
%for the truncation of the disc by the neutron star.

\section*{Acknowledgments}

The X-ray light curves were provided by the ASM/RXTE teams at MIT and at
the RXTE SOF and GOF at NASA's GSFC. This research has made use of NASA's
Astrophysics Data System Bibliographic Services and SIMBAD databases.

\label{lastpage}

\end{document}